\documentclass[aps,prd,10pt,twocolumn,groupedaddress,showpacs,showkeys]{revtex4-1}

\usepackage{amsmath}
\usepackage{amsfonts}
\usepackage{amssymb}
\usepackage{dsfont}
\usepackage{accents}
\usepackage[mathscr]{eucal}
\usepackage{braket}

\usepackage{graphicx}

\newcommand{\rmd}{\ensuremath{\mathrm{d}}}
\newcommand{\rme}{\ensuremath{\mathrm{e}}}
\newcommand{\rmi}{\ensuremath{\mathrm{i}}}

\newcommand{\ub}{\ensuremath{\bar{u}}}
\newcommand{\vb}{\ensuremath{\bar{v}}}

\newcommand{\rf}{\ensuremath{r_{\rm f}}}
\newcommand{\vinf}{\ensuremath{v_\infty}}

\newcommand{\vl}{\ensuremath{v_{\rm l}}}

\newcommand{\apr}{\ensuremath{a_{\rm traj}}}
\newcommand{\arad}{\ensuremath{a_{\rm rad}}}

\newcommand{\Th}{\ensuremath{T_{\rm H}}}

\newcommand{\omegab}{\ensuremath{\bar{\omega}}}

\newcommand{\seff}{\ensuremath{\sigma_{\rm eff}}}

\begin{document}

\title{Quantum frictionless trajectories versus geodesics}

\author{Luis C.\ Barbado}
\email[]{luis.cortes.barbado@univie.ac.at}
\affiliation{Quantenoptik, Quantennanophysik und Quanteninformation, Fakult\"at f\"ur Physik, Universit\"at Wien, Boltzmanngasse 5, 1090 Wien, Austria\\
Instituto de Astrof\'isica de Andaluc\'ia (CSIC), Glorieta de la Astronom\'ia, 18008 Granada, Spain}
\author{Carlos Barcel\'o}
\email[]{carlos@iaa.es}
\affiliation{Instituto de Astrof\'isica de Andaluc\'ia (CSIC), Glorieta de la Astronom\'ia, 18008 Granada, Spain}
\author{Luis J.\ Garay}
\email[]{luisj.garay@ucm.es}
\affiliation{Departamento de F\'{\i}sica Te\'orica II, Universidad Complutense de Madrid, 28040 Madrid, Spain\\
Instituto de Estructura de la Materia (CSIC), Serrano 121, 28006 Madrid, Spain}
\date{\today}

\begin{abstract}
Moving particles outside a star will generally experience quantum friction 
caused by Unruh radiation reaction. There exist however radial trajectories 
that lack this effect (in the outgoing radiation sector, and ignoring 
back-scattering). Along these trajectories, observers perceive 
just stellar emission, without further contribution from the Unruh effect. They 
turn out to have the property that the variations of the Doppler and the 
gravitational shifts compensate each other. They are not geodesics, and their 
proper acceleration obeys an inverse square law, which means that could in 
principle be generated by outgoing stellar radiation. In the case of a black 
hole emitting Hawking radiation, this may lead to a buoyancy scenario. The 
ingoing radiation sector has little effect and seems to slow down the fall even 
further.
\end{abstract}



\maketitle

\section{Introduction}

Inertial trajectories or geodesics is one of the most basic and 
fundamental notions in which modern physics relies. In any flat region of the 
universe (a Minkowskian region), these inertial trajectories are nothing but 
straight lines traveled at constant speeds. In the absence of any external 
force, a body will move along these trajectories: these trajectories are 
frictionless. However, when one considers the quantum nature of the vacuum and 
the unavoidable coupling of material bodies to this vacuum, one could wonder 
whether some quantum friction might appear, making these geodesics not so 
natural.

In Minkowski spacetime, all indicates that this quantum friction 
does not affect inertial trajectories. We can see this by performing a thought 
experiment using a simple Unruh-DeWitt detector~\cite{DeWitt:1980hx} following 
different trajectories. Setting the field initially in the natural Minkowski 
vacuum, only when the detector accelerates it becomes excited due to 
the well known Unruh effect~\cite{Unruh:1976db}, exciting in turn 
the field and so experiencing some quantum friction (or what can be called 
Unruh radiation reaction~\cite{Unruh:1992sw, Parentani:1995iw}). Thus, the 
inertial trajectories in Minkowski spacetime are also quantum frictionless 
trajectories: these two notions coincide.

Here we will show that these two notions become distinct when dealing with 
curved spacetimes. For simplicity and concreteness, we will illustrate 
this issue using radial trajectories in the Schwarzschild spacetime that surrounds any 
spherically symmetric stellar object. We will describe in detail what we call 
\emph{quantum frictionless trajectories,} and also propose a possible buoyancy scenario based on them.

\subsection{Radiation action versus quantum friction}

An Unruh-DeWitt detector is sensitive both to its trajectory and to the state of the quantum fields that it is coupled to. If a detector following an arbitrary trajectory in a curved geometry gets excited, we may wonder whether this excitation is due to the detection of some particle already present in the state of the field, or due to some interaction in which the detector has both perturbed the field and got itself excited. As we will show, this inquiry makes full sense at least in spherically symmetric and asymptotically flat cases, where one can introduce a natural notion of referential vacuum (the Boulware vacuum). Throughout this work we will restrict to these cases (without this structure the above inquiry might loose its content).

We may rephrase the previous inquiry as: \emph{Where does the energy for the excitation come from?} In the first case, the energy for the excitation comes from the energy content of the quantum field, in which case we can say that the detector has detected a particle that already existed in this quantum field. This kind of detection obviously depends on the state of the quantum field, and also on the trajectory of the detector, which modulates the perception of the state. In the second case, the energy for the excitation of the detector can come from the energy supplied by the rockets that make the detector to follow a certain trajectory, or also from the gravitational potential energy of the detector in a curved geometry. In this case, it is clear that the detector has itself created the particle that it is detecting. No such particle would have existed if the detector was not present, since it is the detector that canalizes the energy from other sources towards the quantum field and, through the quantum field, towards itself. This kind of detection is what we identify with a pure Unruh effect.

Again, distinguishing between the two kinds of detection in a general scenario can be difficult or even impossible, but in sherically symmetric and asymptotically flat geometries the distinction can be made in an unambiguous way. In this spacetime, one can put a detector arbitrarily far away from the central stellar object and following a static trajectory, which in that region is also arbitrarily close to an inertial trajectory. If this \emph{far-away-and-static} detector gets excited, then there is no way out: The detector has negligible rockets and is not exploiting its gravitational potential energy, so the energy of the excitation must have been provided by the quantum field. Thus, any detected particle must have been already present in the field, and would have existed even if no detector was there. Unless there are additional sources of radiation, these particles must have been emitted by the stellar object (for example, the emission of light by the Sun, or of Hawking radiation by a black hole), or must be incoming particles from the asymptotic region (for example, the CMB).

Once we have clearly identified the radiation present in the quantum state of the field by using this \emph{far-away-and-static} detector, we can use the equations for the free evolution of the field to propagate it backwards (forwards for ingoing radiation), and see how it ``looked like'' (``will look like'') at any other region of the geometry. Then, for any other detector following an arbitrary trajectory at an arbitrary position, we can distinguish which part of its detection is the radiation present in the state of the field ---that given by the \emph{far-away-and-static} detector adequately propagated---, and which part is due to the perturbation of the field by the detector itself ---the rest of the detection---. We will show how this second part, which we identify with the Unruh effect, only depends on the trajectory of the detector. Moreover, contrary to what one could naively expect, it does not depend directly on the proper acceleration of the trajectory, but rather on its acceleration with respect to the asymptotic region.

Both kinds of detections that we are discussing will in general affect the trajectory of the detector through a scattering process. In the case of the detection of particles already present in the quantum field, this process is just the collision of two objects: the detected particle and the detector itself. Its effect on the trajectory of the detector is what we shall call \emph{radiation action.} In the case of the Unruh effect, the detector both gets excited and excites in turn the field, emitting what is usually called \emph{Unruh radiation.} This emission affects the trajectory of the detector through back-reaction, and this back-reaction is what we shall call \emph{quantum friction.}

\section{Frictionless trajectories}

To expose our arguments in the simplest possible terms, let us consider an Unruh-DeWitt detector coupled to the $s$-wave sector of a Klein-Gordon massless real field~$\phi$, and moving along radial trajectories in the Schwarzschild exterior geometry. By ignoring the back-scattering of the radiation on the geometry we obtain an effective conformal field theory with a Klein-Gordon equation equivalent to that in two-dimensional flat spacetime: 
\begin{equation}
\square \phi=	\frac{\partial}{\partial \ub} \frac{\partial}{\partial \vb} \phi = 0,
\label{klein-gordon}
\end{equation}
where~$\ub$, $\vb$ are the Eddington-Finkelstein null coordinates. Conformal invariance makes the ingoing and outgoing radiation sectors decouple, so that the general solution is of the form $\phi = \phi (\vb) + \phi (\ub)$.

\subsection{Definition of the trajectories}

We want to calculate the radial trajectories that lack quantum friction with the field under consideration. With this aim, we will work in an scenario in which we already get rid of the radiation-action part, that is, we will work with a state of the quantum field without radiation sources. (In other words, we will fix the state such that the \emph{far-away-and-static} detector does not get excited.) By getting rid of the radiation action, no quantum friction will be automatically equivalent to no radiation detection at all. We can then compute the quantum frictionless trajectories in an easier way: just by computing the trajectories with no radiation detection in that particular state. In the Schwarzschild exterior geometry, the state without radiation sources is the Boulware vacuum~\cite{Boulware:1974dm}. Thus, in order to compute the quantum frictionless trajectories we will work in the Boulware vacuum state. What happens along these trajectories in other states will be discussed later on.

It is already known that, in the Boulware vacuum, Unruh-DeWitt detectors 
experience no detection along any static trajectory (see, e.g., 
\cite{Louko:2007mu, Hodgkinson:2013tsa}; this problem has an analogue in 
classical electrodynamics, see \cite{Fulton1960499, kovetz}). But, are
there any other trajectories for which Unruh-DeWitt 
detectors will not get excited? The short answer is that, 
strictly speaking, there are no more trajectories with zero excitation. However, we will see 
that, under the already taken assumption of neglecting back-scattering, and limiting ourselves to one radial radiation sector, a non-trivial notion of quantum frictionless trajectory emerges. We will first focus on the outgoing radiation sector~$\phi (\ub)$, paying attention to the ingoing sector later on.

Let the Unruh-DeWitt detector follow a radial trajectory~$r(\tau)$, with~$\tau$ being the proper time, from which 
one can compute~$\ub(\tau)$ using the Schwarzschild metric. In the Boulware vacuum state, the probability amplitude for the detector 
to get excited from its ground state to its excited state (with an energy 
gap~$\omega$), exciting in turn some mode~$\rme^{-\rmi \omegab \ub}$ of the 
field, is proportional to \cite{Birrell:1982ix}
\begin{equation}
	\int_{-\infty}^\infty \rmd \tau\ \rme^{-\rmi \left[ \omega \tau + \omegab \ub (\tau) \right]}.
\label{amplitude}
\end{equation}
This quantity is also proportional to the Bogoliubov 
coefficient~$\beta_{\omega, \omegab}$ between the modes associated with the 
vacuum state and the natural modes associated with the detector's
trajectory~\cite{Birrell:1982ix}. We could have used a wave packet basis to 
compute the local probability amplitude around some point of the trajectory, 
but for simplicity we will only consider global trajectories. The 
condition on the trajectory~$\ub(\tau)$ such that the detector perceives no radiation, which in the Boulware vacuum is equivalent to the definition of quantum frictionless trajectory, amounts to the vanishing of the probability amplitude~(\ref{amplitude}) regardless of the values of~$\omega$ and~$\omegab$. One can easily check that this condition 
entails that the null coordinate~$\ub$ must have a linear dependence in~$\tau$:
\begin{equation}
	\dot{\ub} = \dot{t} - \frac{1}{1-2M / r} \dot{r} = C,
\label{differential_equation}
\end{equation}
where~$C$ is a positive constant and the dot denotes derivative with respect to $\tau$. We can simplify this differential equation using the Schwarzschild metric:
\begin{equation}
	-\frac{M}{\sqrt{1-q}}\ \rmd \tau = \frac{r}{q r/(2M) - 1}\ \rmd r,
\label{differential_equation_simpl}
\end{equation}
where~$q := 1-1/C^2$. Equation~(\ref{differential_equation_simpl}) is the differential equation characterizing the quantum frictionless trajectories. Let us analyze the explicit form of the solutions.

\subsection{Classification of the trajectories}

The solutions to equation~(\ref{differential_equation_simpl}) fall into three different types, depending on the value of~$q$ ---in all the cases we have arbitrarily fixed the origin of proper time---:

\begin{enumerate}
	\item
		$0 < q < 1$ corresponds to the solution:
		\begin{equation}
			r(\tau) = \frac{2M}{q} \left[ 1 + W_0 \left(\epsilon\ \rme^{- g \tau} \right) \right],
		\label{trajectory}
		\end{equation}
		where $g := q^2/(4 M \sqrt{1-q})$, $W_0(z)$ is the branch of the Lambert~$W$ function with~$W_0(z) \in \mathds{R}$ and~$W_0 (z) \geq -1$ for~$z \in [-1/\rme, \infty)$, and $\epsilon$~is a discrete parameter further classifying the trajectories. These trajectories reach an asymptotic radius~$\rf := 2M/q$ with asymptotic proper acceleration~$g$. The discrete parameter $\epsilon$ determines the side from where the asymptotic radius is approached: 
		
		\begin{enumerate}
			\item
				$\epsilon = 1$ corresponds to ingoing trajectories from the asymptotic region with initial radial velocity~$\vinf = q/(q-2)$ to~$\rf$ (FIG.~\ref{fig_ingoing}),
				
			\item
				$\epsilon = 0$ represents an static trajectory at~$r = \rf$,
				
			\item
				$\epsilon = -1$ corresponds to outgoing trajectories from the surface of the stellar object to $\rf$. 
				
		\end{enumerate}
		
		%------------------------------------------------------------------------------------------------------------------------------------------
		\begin{figure}[h]
		\includegraphics[width=\columnwidth]{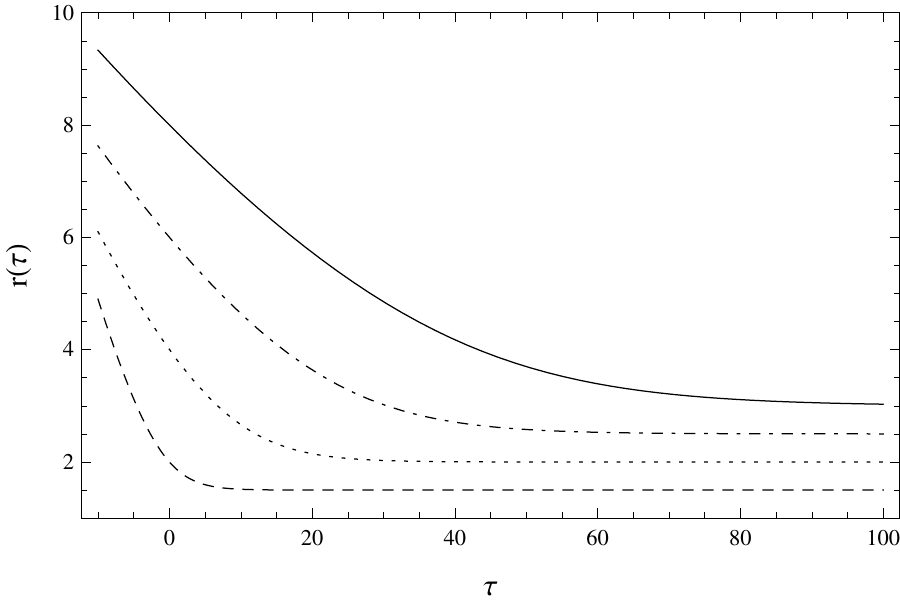}
		\caption{Ingoing trajectories~$r (\tau)$ in~(\ref{trajectory}) for~$\epsilon = 1$ and $\rf = (3M,\ 4M,\ 5M,\ 6M)$ (dashed, dotted, dash-dot and solid line, respectively). We use~$2M = 1$ units. The origin of proper time has been shifted differently for each trajectory to allow for a clearer presentation.\label{fig_ingoing}}
		\end{figure}
		%------------------------------------------------------------------------------------------------------------------------------------------
		
	\item
		$q = 0$ corresponds to an outgoing trajectory escaping to the asymptotic region with zero radial velocity:
		\begin{equation}
			r(\tau) = \sqrt{2M \tau}.
		\label{trajectory_limit}
		\end{equation}

	\item
		$q < 0$ corresponds to outgoing trajectories escaping to the asymptotic region with radial velocity~$\vinf = q/(q-2)$:
		\begin{equation}
			r(\tau) = \frac{2M}{q} \left[ 1 + W_{-1} \left(- \rme^{- g \tau} \right) \right],
		\label{trajectory_escape}
		\end{equation}
		where~$W_{-1}(z)$ is the branch of the Lambert~$W$ function with~$W_{-1} (z) \in \mathds{R}$ and~$W_{-1} (z) \leq -1$ for~$z \in [-1/\rme, 0)$.

\end{enumerate}
In the cases~1(c), 2 and~3, the proper time range is bounded at the past 
by the condition~$r(\tau) > 2M$.

\subsection{Physical interpretation}

One can notice that, since the Boulware vacuum state is fully determined by the geometry, the differential equation~(\ref{differential_equation}) is also purely geometrical. Here we shall stress one point: The fact that the Boulware vacuum fixes the quantum frictionless trajectories through the previous calculation does not mean that there would be other quantum frictionless trajectories for other states. The only special characteristic of the Boulware vacuum is that it lacks radiation sources, so that in this state no quantum friction is equivalent to no radiation detection. But, once computed in one way or another, the set of quantum frictionless trajectories is unique. There are no different sets of trajectories that one should follow to avoid quantum friction depending on the state.

There is a clear physical interpretation for equation~(\ref{differential_equation}): An observer following a trajectory described by this equation will keep his clock rate ---or frequency shift--- constant with respect to the asymptotic region, when compared by emitting outgoing light pulses. In fact, there is a very intuitive way to understand how observers maintain this constant relation. One can write equation~(\ref{differential_equation}) as
\begin{equation}
	\dot{ \ub} = \sqrt{\frac{1-\vl}{1+\vl}} \frac{1}{\sqrt{1-\frac{2M}{r}}} = C,
\label{compensation}
\end{equation}
where~$\vl := \dot{r} / \sqrt{1- 2M/r + \dot{r}^2}$ is the local velocity of the trajectory with respect to the static geometry~\cite{Barbado:2012pt} ---and thus both with respect to the asymptotic region and to the stellar object---. We see that this expression is the product of two factors with direct physical meaning: the former is a Doppler shift factor and the latter a gravitational blueshift factor. Written in this way, it is clear that the frictionless condition~(\ref{differential_equation}) imposes that the product of the two clock effects on frequency shifts must be constant, and thus the variation of one must be compensated by an opposite variation of the other. 

As an example, let us consider the ingoing trajectories~1(a). They start at the 
asymptotic region with some ingoing velocity, that is, some Doppler blueshift for the 
outgoing radiation. Along the trajectory, the gravitational blueshift starts 
to grow and the velocity of the trajectory to diminish, lowering the Doppler 
blueshift and so maintaining the product of factors constant. At the end, the 
trajectory tends asymptotically to a constant radius where the final gravitational blueshift exactly replaces the 
initial Doppler blueshift. Analogous arguments apply to the other 
trajectories.

The quantum frictionless trajectories are nowhere geodesic, although they approach geodesics at the asymptotic region. Thus, observers following these trajectories experience some proper acceleration, while they experience no quantum friction. On the other hand, geodesic observers do experience quantum friction because they accelerate with respect to the asymptotic region, while having no proper acceleration (by definition). We can say that the quantum friction is due to the ``rubbing'' of any device with the ``Boulwarean part'' of the quantum vacuum. And it is the change in the clock rate with respect to the asymptotic region that determines this friction, independently of the cause of this acceleration, whether gravitational or not.

In any other field state different from the Boulware vacuum, an observer 
following a quantum frictionless trajectory will only perceive the radiation emitted 
by the stellar object to the asymptotic region, just shifted with the constant 
frequency shift~$C$ characteristic of the trajectory. And nothing more. This can be easily seen from the linear relation~$\ub = C \tau$ that defines the trajectories: The natural modes for the observers following the trajectories~$\rme^{-\rmi \omega \tau}$ are just the natural modes in the asymptotic region~$\rme^{-\rmi (\omega/C) \ub}$ with the constant frequency shift. As we already argued, this perceived radiation will in general affect the motion of a test object following the frictionless trajectory through a scattering process. We will precisely use this fact in the buoyancy scenario presented later on. But this effect cannot be 
considered as ``friction'' or ``back-reaction'', since it is not the test-object motion 
that is producing the radiation: It is just the expected \emph{radiation 
action} of the stellar emission on the test object.

Notice also that, contrary to the notion of geodesic, the quantum 
frictionless trajectories seem to have non-local properties: 
An observer needs to know where he is in the geometry and with 
what velocity with respect to the asymptotic region. However, this 
non-locality is just apparent. The asymptotic reference is necessary to define 
what one means by ``Boulwarean part'' of the quantum vacuum. Once 
this is fixed, an observer can adjust its rockets to follow a quantum 
frictionless trajectory by just looking at the local properties of this vacuum 
structure.

%-----------------------------------------------------------------------------------------------

\section{Ingoing radiation sector}

It is easy to check that the trajectories lacking quantum friction in the ingoing radiation sector are actually the same frictionless trajectories in the outgoing radiation sector but followed \emph{in the opposite direction:} In equation~(\ref{compensation}) one should just flip the sign of the velocity. Thus, except for the static trajectories~1(b), the frictionless trajectories in the outgoing sector will experience some quantum friction in the ingoing sector, and vice versa.

Given a frictionless trajectory in the outgoing radiation sector, one could calculate how much it deviates from being frictionless in the ingoing sector, for example by calculating the corresponding Bogoliubov coefficients in this sector. In general, this calculation has to be done numerically. However, we can estimate the amount of excitation of the Unruh-DeWitt detector in this sector by using the so-called \emph{effective temperature function}~$\kappa(\tau):=-\ddot{ \vb} /\dot{ \vb}$. This function was introduced in~\cite{Barcelo:2010pj, Barcelo:2010xk} and extensively used in~\cite{Barbado:2011dx, Barbado:2012pt, Smerlak:2013sga, Smerlak:2013cha} to estimate the radiation perception of a detector in different trajectories and different vacua in black hole spacetimes. The absolute value of this function is, under certain adiabatic condition, $2 \pi$ times the temperature perceived by an observer at time~$\tau$. When the function deviates from the adiabatic condition, one can still use it as an estimator of the power of the radiation perceived, only having in mind that there will be other non-thermal contributions to the spectrum~\cite{Barbado:2012fy}.

\begin{figure}[h]
\includegraphics[width=\columnwidth]{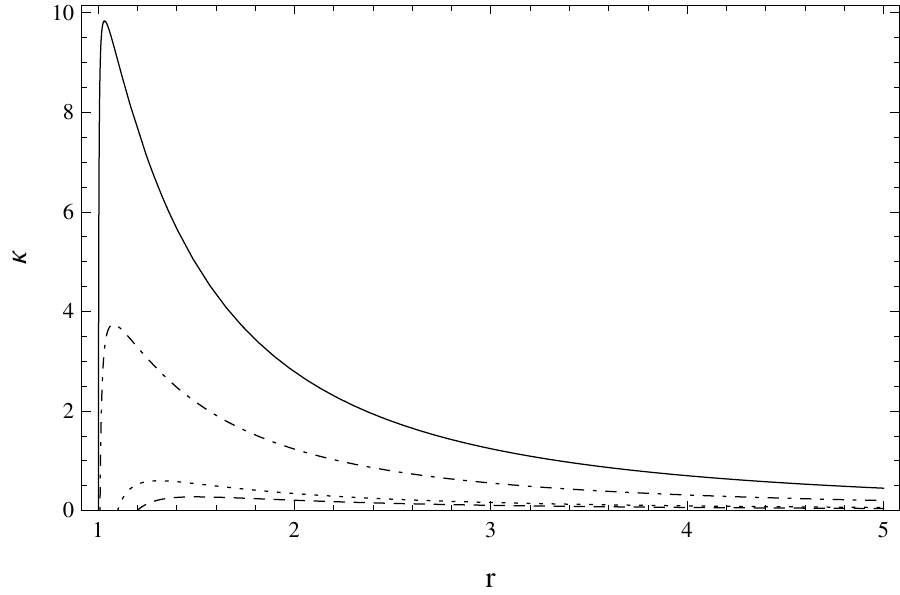}
\caption{Absolute value of the effective temperature function associated with the quantum friction in the ingoing sector for different trajectories~1(a) with $\rf = (2.4M,\ 2.2M,\ 2.02M,\ 2.004M)$ (dashed, dotted, dash-dot and solid line, respectively). We use~$2M = 1$ units.\label{fig_kappa}}
\end{figure}

We have computed the effective temperature function of the ingoing radiation sector for the quantum frictionless trajectories~1(a) ---the most physically interesting, see next section--- in the Boulware vacuum (again, to avoid any external source of radiation). As a function of the radial position, its absolute value reads:
\begin{equation}
|\kappa (r)| = \frac{M}{r^2 \sqrt{1-2M/\rf}} \left( 1- \frac{1-2M/\rf}{1-2M/r} \right).
\label{kappa_ingoing}
\end{equation}
We plot this quantity for some sample trajectories in~FIG.~\ref{fig_kappa}. The absolute value of the effective temperature vanishes uniformly when~$\rf \to \infty$, while it diverges non-uniformly when~$\rf \to 2M$. When computed along an specific trajectory characterized by $\rf$, $\kappa$ vanishes in the asymptotic past and future, while it presents a single peak at~$r \lesssim 1.5 \rf$ for~$\rf \gg 2M$ and at~$r \gtrsim \rf$ for~$\rf \gtrsim 2M$.

%------------------------------------------------------------------------------------------------------------------------------------------
%------------------------------------------------------------------------------------------------------------------------------------------

In conclusion, strictly speaking there 
are no frictionless trajectories in a Schwarzschild background except for the 
static ones. However, in the next section we will argue that the approximate 
notion we have presented here could have important applications in black hole 
physics. 

%-----------------------------------------------------------------------------------------------

\section{Buoyancy processes}

Let us consider an scenario in which an stellar object emits constant and 
isotropic radiation of the quantum field under consideration. A black hole emitting Hawking 
radiation~\cite{Hawking:1974sw} is a paradigmatic case, but our assertions are more general and do not depend on the spectrum or on the nature of the 
emitting body. We will show that under the previous circumstances 
the radiation pressure of the emitted quanta could cause a test object falling 
radially to follow the trajectories~1(a), instead of 
a geodesic. This could lead to a self-consistent buoyancy scenario, in which 
the test object ends up floating at constant radius.

The radial proper acceleration needed to follow a quantum frictionless trajectory can be computed from~(\ref{differential_equation}), yielding
\begin{equation}
\apr = \frac{CM}{r^2}.
\label{acceleration_traj}
\end{equation}
Now let us consider that the total 
power emitted by the stellar object as measured in the asymptotic 
region is~$P$. The irradiance ---power per unit area--- perceived by an observer at a radius~$r$ and moving with local radial velocity~$\vl$ is
\begin{equation}
S = \frac{1-\vl}{1+\vl} \frac{1}{1 - \frac{2M}{r}} \frac{P}{4 \pi r^2},
\label{irradiance}
\end{equation}
where the first two factors are the expected Doppler and gravitational shifts (note that the power scales 
as~$({\rm time})^{-2}$). For a test object moving along a quantum frictionless trajectory the product of the first two factors is equal to~$C^2$  ---see equation~(\ref{compensation})---, 
and therefore $S = C^2 P / (4 \pi r^2)$. Considering a test object 
with rest mass~$m$ and effective cross section~$\seff$, and considering elastic 
scattering of the radiation, the proper acceleration produced by the radiation 
pressure over the object would be
\begin{equation}
\arad = \frac{2 \seff S}{m} = \frac{\seff C^2 P}{2 \pi m r^2}.
\label{acceleration_rad}
\end{equation}

Since the functional dependence of~$\arad$ with~$r$ is the same as that of~$\apr$ in~(\ref{acceleration_traj}), if we fix the quotient~$\seff/m$ to
\begin{equation}
\frac{\seff}{m} = \frac{2 \pi M}{C P},
\label{quotient}
\end{equation}
we get $\apr = \arad$ all along the trajectory. Thus, the radiation emitted by the stellar object (which is the only radiation that the object perceives along a quantum frictionless trajectory) would be precisely the source of proper acceleration needed for the object to follow a quantum frictionless trajectory. For the trajectories~1(a), this would lead to a self-consistent buoyancy scenario.

A particularly important example of 
constant and isotropic radiation is that expected from a black hole 
with mass~$M$. Its temperature is $\Th = 1/(8 \pi M)$ and its 
total power $P = 16 \pi M^2 \sigma \Th^4$, with~$\sigma$ being the Stefan-Boltzmann constant. Although this power is tiny for any astrophysical black hole, the irradiance detected by a test object can be 
significant if it is magnified by a large frequency shift. This would be the case, for 
instance, for the quantum frictionless trajectories~1(a) that tend to static 
positions very close to the event horizon. Along these trajectories, the falling object would 
detect outgoing radiation at constant temperature equal to the Hawking temperature 
multiplied by the blueshift factor~$C=1/\sqrt{1-2M/\rf}$ associated to its final radial position. 
Thus, this potentially strong radiation pressure could provide the 
force needed to keep the infalling object along the quantum 
frictionless trajectory. However, note that for achieving this buoyancy 
scenario one would need in principle to ``shoot'' the test object from 
the asymptotic region towards the black hole with a Doppler blueshift 
equivalent to the final gravitational blueshift.

In this description we have not included the 
effect of the non-zero quantum friction in the ingoing sector.
In general, its relative importance with respect to the radiation pressure responsible for the buoyancy will vary along the 
trajectory, and will depend both on the particular trajectory 
and on the total emission of the stellar source. However, it will not depend on the 
test-object cross section or mass, at least at first order: Larger cross 
section means both more radiation detection and more quantum 
friction; and larger mass means both less deviation due to radiation pressure 
and due to quantum friction. Moreover, any friction in the ingoing sector would 
tend to slow down even further the velocity of the infalling object, in 
principle helping to attain a fixed radius buoyant trajectory.

\begin{figure}[h]
\includegraphics[width=\columnwidth]{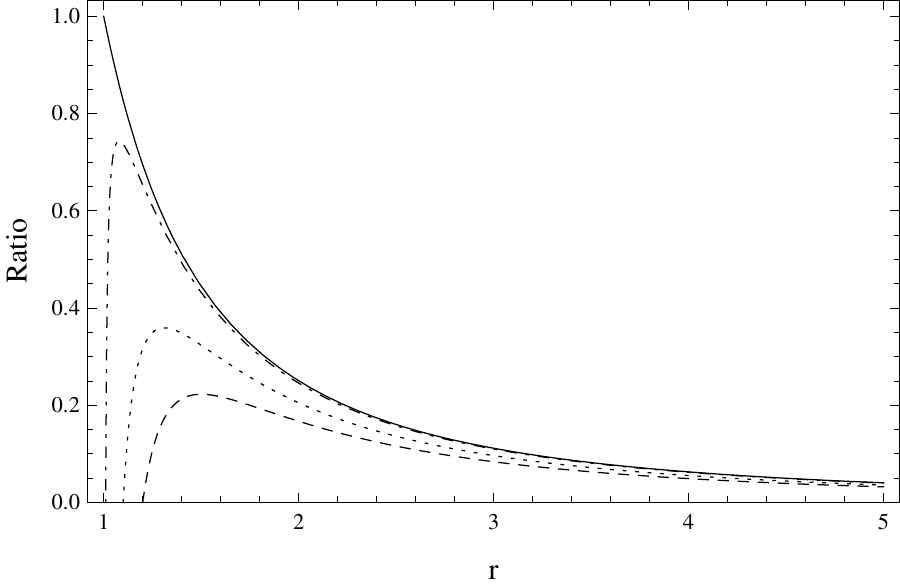}
\caption{Ratio between the absolute value of the effective temperature function associated with the quantum friction in the ingoing sector and that associated with the perception of the outgoing Hawking radiation along the trajectory, for different trajectories~1(a) with $\rf = (2.4M,\ 2.2M,\ 2.02M,\ 2M)$ (dashed, dotted, dash-dot and solid line, respectively). We use~$2M = 1$ units.\label{fig_ratios}}
\end{figure}

A first and direct evaluation of this relative importance would be to compare the effective temperature functions associated with the ingoing and outgoing radiation sectors. As an example, we have compared the two effective temperature functions in the case in which the source for the buoyancy is the Hawking radiation from a black hole. That is, we have computed the ratio between the quantity in~(\ref{kappa_ingoing}), and the effective temperature function associated with the perception of the outgoing Hawking radiation along the trajectory, equal to~$2 \pi C \Th = 1/(4M \sqrt{1-2M/\rf})$. This ratio reads:
\begin{equation}
\frac{|\kappa (r)|}{2 \pi C \Th} = \frac{4 M^2}{r^2} \left( 1 - \frac{1-2M/\rf}{1-2M/r} \right).
\label{ratio}
\end{equation}
By analyzing this quantity, we can see that the absolute value of the effective temperature associated with the quantum friction is usually much lower than that associated with the Hawking radiation (see~FIG.~\ref{fig_ratios}). It tends to be at most equal only when the asymptotic radius~$\rf$ is arbitrarily close to the event horizon, and only just before the trajectory approaches this radius.

%-----------------------------------------------------------------------------------------------

\section{Final remarks}

In this work we have introduced the notion of quantum frictionless trajectories as a distinct notion from that of geodesics. Although there are no strict quantum frictionless trajectories beyond the static trajectories, the notion appears well defined when separating the effect into the ingoing and outgoing sectors and neglecting back-scattering. We think that, even if the quantum frictionless notion and the related buoyancy character of some of the frictionless trajectories are only approximate, in practice they constitute poignant notions to better appreciate the distinction between acceleration with respect to the vacuum or with respect to the spacetime. For instance, it would be very interesting to probe this distinction in analogue gravity experiments. Is the buoyancy scenario implementable in a realistic analogue setup? We leave this question for future work. 

%------------------------------------------------------------------------------------------------------------------------------------------
%------------------------------------------------------------------------------------------------------------------------------------------

%-----------------------------------------------------------------------------------------------

\begin{acknowledgments}
The authors wish to thank two anonymous referees for their constructive criticism, 
which helped to improve this article.
Financial support was provided by the Spanish MICINN through Projects No. FIS2011-30145-C03-01 and FIS2011-30145-C03-02 (with FEDER contribution), and by the Junta de Andaluc\'ia
through Project No. FQM219.
\end{acknowledgments}

%-----------------------------------------------------------------------------------------------

\bibliography{biblio}

\end{document}